# Simulation of Intravoxel Incoherent Perfusion Signal Using a Realistic Capillary Network of a Mouse Brain

Valerie Phi van[1], Franca Schmid[2,3], Georg Spinner[4], Sebastian Kozerke[4], Christian Federau[4,*]

[1]*University Hospital Zürich, Zürich, Switzerland*

[2]*Institute of Pharmacology and Toxicology, University of Zurich, Zürich, Switzerland*

[3]*Institute of Fluid Dynamics, ETH Zurich, Zurich, Switzerland*

[4]*Institute for Biomedical Engineering, ETH and University of Zürich, Zürich, Switzerland*

*Corresponding Author:*
Christian Federau
University and ETH Zürich
Institute for Biomedical Engineering
Gloriastrasse 35
8092 Zürich
Switzerland
Email: federau@biomed.ee.ethz.ch

Purpose

To simulate the intravoxel incoherent perfusion magnetic resonance magnitude signal from the motion of blood particles in three realistic vascular network graphs from a mouse brain.

Methods

In three networks generated from the cortex of a mouse scanned by two-photon laser microscopy, blood flow in each vessel was simulated using Poiseuille's law. The trajectories, flow speeds and phases acquired by a fixed number of simulated blood particles during a Stejskal-Tanner monopolar pulse gradient scheme were computed. The resulting magnitude signal as a function of b-value was obtained by integrating all phases and the pseudo-diffusion coefficient D* was estimated by fitting an exponential signal decay. To better understand the anatomical source of the IVIM perfusion signal, the above was repeated by restricting the simulation to various types of vessels.

Results

The characteristics of the three microvascular networks were respectively: vessel lengths [mean±std. dev.]: 67.2±53.6µm, 59.8±46.2µm, and 64.5±50.9µm; diameters: 6.0±3.5µm, network 2: 5.7±3.6µm, and network 3: 6.1±3.7µm; simulated blood velocity: 0.9±1.7µm/ms, 1.4±2.5µm/ms and 0.7±2.1µm/ms. Exponential fitting of the simulated signal decay as a function of b-value resulted in the following D* [$10^{-3}$ mm$^2$/s]: 31.7, 40.4 and 33.4. The signal decay for low b-values was the largest in the larger vessels, but the smaller vessels and the capillaries accounted more to the total volume of the networks.

Conclusion

This simulation improves the theoretical understanding of the IVIM perfusion estimation method by directly linking the MR IVIM perfusion signal to an ultra-high resolution measurement of the microvascular network and a realistic blood flow simulation.

**INTRODUCTION**

Measuring specifically microvascular perfusion is of particular clinical interest, because microvascular dysfunction plays an important role in many important diseases, such as vascular dementia, lacunar infarcts, or diabetes. For example microvascular complications in patients with diabetes are the cause of blindness, renal failure non-traumatic amputations, and a powerful predictors of cardiovascular complications.[1]

Microvascular perfusion can be assessed with Intravoxel Incoherent Motion (IVIM) magnetic resonance (MR) perfusion imaging without injection of intravenous contrast agent.[2] The method has seen a regain of research interest in the last decade, both in the brain[3] and in the body,[4] and many recent experimental and clinical studies have demonstrated the feasibility to measure microvascular perfusion with IVIM.[5] For example, IVIM perfusion has been shown to be altered early in patients with diabetes[6] and the dependence on the local microvascular perfusion has been used to assess the quality of collateral blood flow in the context of stroke[7] or to detect vasospasm at the microvascular level after aneurysm rupture.[8] But the microscopic anatomical origin of the Intravoxel Incoherent Motion (IVIM) perfusion signal is not well understood. It is assumed to arise from motion of the blood inside the microvasculature, but an ultimate experimental validation of this assumption is lacking. In addition, the exact relationships between the IVIM perfusion parameters, the micro-vascular network structure, and the blood flow, are not clearly established.

The purpose of this study was to simulate blood flow in three realistic microvascular networks, which were imaged at ultrahigh resolution, and to derive the expected IVIM magnetic resonance magnitude signal and from that, the expected IVIM pseudo-diffusion coefficient $D^*$.

The microvascular networks, embedded in a tissue volume of 1mm$^3$, were obtained by two-photon laser scanning microscopy and contained on average 15'000 vessels.[9] Realistic pressure boundary conditions at all in- and outflows were assigned, and Poiseuille's law was used to compute flow rate in each vessel.[10] The trajectories of a fixed number of blood particles were then computed for a defined time interval, and used to calculate the phase acquired by each blood proton during standard Stejskal-Tanner monopolar pulse gradient scheme as used in a typical IVIM experiment.[11] The sum of all phases resulted in the simulated MR magnitude signal as a function of the b-value, and the IVIM pseudo-diffusion coefficient D* was obtained by fitting an exponential signal decay. Finally, to better understand the anatomical source of the IVIM effect, we decomposed the signal by restricting the simulation to the various types of vessels: artery, arterioles, capillary bed, venules and veins.

**METHODS**

*Microvascular Networks*

The three microvascular networks were acquired by two-photon laser scanning microscopy[9]. Each network is embedded in a tissue volume of approximately 1 mm$^3$ and contains on average 15'000 vessels. An in-depth description of the pre-processing steps for the blood flow simulations is available here.[10] In brief, the microvascular networks are stored in a graph format, i.e. they consist of a set of nodes (bifurcations), which are connected by edges (vessels). The tortuosity of the vessels is stored in coordinates as edge attributes. Further edge attributes are the effective vessel diameter, the vessel length and the vessel type.

*Decomposition into vessel types*

The vessel type was assigned by following individual penetrating vessels from the cortical surface and by applying a diameter criterion: If two consecutive vessel branches had a diameter < 7µm, the second branch and the following were considered to be part of the capillary bed. To obtain a capillary diameter distribution consistent with literature data,[12,13] a histogram-based up-scaling approach was applied, which was based on a beta distribution with a mean of 4 µm and a standard deviation of 1 µm. A three dimensional representation of the microvascular networks under investigation is depicted in **Figure 1**.

*Blood Flow Simulation*

In the following, a brief summary of the key aspects of the blood flow model is provided. A more detailed description is available here.[10] The blood flow model is based on the small Reynolds number in the microvasculature, which allows the use of Poiseuille's law to compute the flow rate between vertices, i.e.

$$q_{ij} = \frac{\pi d_{ij}^4 (p_i - p_j)}{128 \mu^{eff} L_{ij}}$$

where $q_{ij}$ is the flow rate in the vessel connecting node $i$ and $j$, $d_{ij}$ and $L_{ij}$ are the vessel diameter and length, respectively. The variables $p_i$ and $p_j$ are the pressure at node $i$ and $j$ respectively, and $\mu^{eff}$ is the effective viscosity within the vessel. The effective viscosity is a function of the vessel diameter and the local red blood cell (RBC) density, i.e. the hematocrit, and is computed based on empirical equations.[14] To compute the local hematocrit, we track individual RBCs through the microvascular network. Here, we account for Fahraeus-Lindqvist effect, Fahraeus effect and phase separation.[14–16] In non-capillaries, the fractional RBC flow at divergent bifurcations is obtained from empirical equations.[14] At capillary bifurcations, the RBCs follow the path of the largest pressure force. By assigning pressure boundary conditions and a physiological inflow hematocrit of 0.3[17] a well-posed problem is obtained, which together with the continuity equation can be transformed into a linear system of equations. By solving the linear system of equations, the pressure value at each vertex can be computed. Based on the pressure field, the blood flow rate, the flow velocity and the flow direction can be obtained for each vessel. Subsequently, the RBCs are propagated for the time step $\Delta t$ and the vessel resistances are updated based on the novel RBC distribution. This procedure is repeated until a statistical steady state is obtained. Due to the fluctuating RBC distribution the flow field is non-steady and hence, time-averaged simulation results are used for further analysis.

*Particle Tracking*

The trajectories of a fixed number of 20'000 particles were computed along the network architecture, using the simulated characteristic flow speed and pressure gradient for a defined time interval, which is the diffusion time of the Stejskal-Tanner monopolar pulse gradient scheme (see below). The particle track was created as follow:

1.) The particle was placed in a vessel chosen randomly. The starting position inside this vessel itself was chosen randomly as well, according to the relative volume of the vessel segment compared to the volume of the full vessel.

2.) The velocity of the particle in this vessel segment was calculated by the flow within the vessel and its diameter.

3.) The particle followed the tortuosity of the given vessel according to the tortuosity coordinates with the given velocity until it reaches a branching point, at which it entered a new vessel with lower pressure gradient according to the relative flow of all vessels at this branching point.

4.) The steps 2 and 3 were repeated until the defined time interval was reached.

5.) If the particle left the network during the defined time interval, the particle was not considered in the signal simulation.

*IVIM MR Magnitude Signal Simulation*

A Stejskal-Tanner monopolar pulse gradient scheme was used for the simulation, with following parameters[11]: b-values: 0-800 s/mm²; fixed diffusion time: Δ = 100 ms; fixed gradient time: δ = 50 ms. The used gradient strength was calculated for 100 b-values between 0-800 s/mm² using following formula: $b = e^{ax} - 1$, with a = 0.0621661 and x = [0,1,2....100] giving a higher b-value density near b=0 s/mm² and a lower density for the larger b-values, near b = 800 s/mm².

The gradient ramps were neglected.

The track of each particle $n$ was divided into parts $i$ in each vessel segment with constant velocity and constant gradient $g$. The accumulated phase acquired by this particle was

$$\phi_{n,k} = \gamma \int_{t_0}^{t_{end}} r(t) \cdot g \, dt = \gamma \sum_i \left[ r_{0,i} \cdot g \cdot (t_{i,end} - t_{i,0}) + \frac{1}{2} v_i \cdot g \cdot (t_{i,end}^2 - t_{i,0}^2) \right]$$

where $\gamma$ is the gyromagnetic ratio of a proton, $r$ the position, $v$ the velocity, and $t$ the time. The signal amplitude over all $N$ particles was

$$S = \frac{1}{N}\left|\sum_n e^{i\phi_n}\right|$$

The gradient strengths of the Stejskal-Tanner monopolar standard sequence were calculated according to following equation:[18]

$$|g| = \frac{\sqrt{b}}{\gamma \Delta^{3/2} \sqrt{F}}$$

where $F$ is 12 for the Stejskal-Tanner sequence and $\Delta$ is the diffusion time.

*IVIM MR Signal Fit*

The trace of the MR signal was calculated with an applied gradient in x-, y and z-direction. A mono-exponential fit of the trace of the signal was used to obtain the pseudo-diffusion coefficient D*:

$$S = e^{-bD^*}$$

*Restriction of the Simulation to the Various Anatomical Vessel Types*

To better understand the anatomical source of the IVIM perfusion signal, the above was repeated by restricting the simulation to the various types of vessels: capillary, arterioles, venules and pial vessels. This was done by ignoring all of the particles outside the vessel under consideration, as well as by ignoring all of the particles leaving the vessel type under consideration in the signal simulation.

*Statistics*

Statistical analysis was performed using MATLAB version 2015a (MathWorks, Natick, MA). Statistical significance was assessed using Student's t-test with a significance level α of 0.05.

**RESULTS**

*Network Characteristics*:

All three networks differed in their structures. Mean capillary segment lengths of the extracted capillary graphs was for network 1: [mean ± std. dev.] 67.2 ± 53.6 µm, network 2: 59.8 ± 46.2 µm, and network 3: 64.5 ± 50.9 µm. The mean of the diameter distribution was for network 1: 6.0 ± 3.5 µm, network 2: 5.7 ± 3.6 µm, and network 3: 6.1 ± 3.7 µm (**Figure 2**). Most capillary branches in all three networks were connected to 3 branches, with a maximum of 6 connections (**Figure 3**).

*Blood Flow Simulation*

The mean simulated blood speed was 0.9 ± 1.7 µm/ms in network 1, 1.4 ± 2.5 µm/ms in network 2 and 0.7 ± 2.1 µm/ms in network 3 (**Figure 2**). Arteries showed, as expected, significantly higher blood velocities compared to capillaries (12.5 ± 10.2 versus 0.7 ± 1.1 µm/ms, $p < 0.05$), see **Figure 3**.

*MR Signal Simulation*

The IVIM MR signal decay as a function of b-value was faster than mono-exponential for all three capillary networks **(Figure 4)**. Exponential fitting resulted in the following IVIM pseudo diffusion coefficient D*: 31.7 x $10^{-3}$ mm$^2$/s (network 1), 40.4 x $10^{-3}$ mm$^2$/s (network 2) and 33.4 x $10^{-3}$ mm$^2$/s (network 3).

*Dependence of the IVIM MR Signal on the Vessel Type*

The signal decay was strongly dependent on the type of vessels considered **(Figure 5, Table 1)**. Slowest signal decay for low b-values was seen in the capillary network (D* range for all three networks: 6.3 - 9.8 x $10^{-3}$ mm$^2$/s), a moderate decay for the descending arterioles (21.7 – 40.4 x $10^{-3}$ mm$^2$/s) and veins (61.9 – 219.3 x $10^{-3}$ mm$^2$/s), and a steep signal decay for the pial artery (172.1 – 833.3 x $10^{-3}$ mm$^2$/s). The contribution from the capillaries to the total volume of the network was the largest of all vessel types.

**DISCUSSION**

We simulated blood motion in three realistic microvascular networks obtained by two-photon laser microscopy in the mouse brain and computed the effect of this motion on the MR IVIM perfusion signal for b-values between 0 and 800 s/mm$^2$. Our findings are in good agreement with the assumption of a microvascular source of the IVIM perfusion signal, although our simulation derived pseudo-diffusion coefficient D* values were in the upper range compared with in-vivo measured signal. The pseudo-diffusion coefficient D* has been reported[11,19,20] between 7 x $10^{-3}$ mm$^2$/s and 17 x $10^{-3}$ mm$^2$/s for young healthy adults, but D* of 31 x $10^{-3}$ mm$^2$/s have been reported in the gray matter after functional activation,[20] and between 28.49 and 73.96 x $10^{-3}$ mm$^2$/s in region of interests drawn in pathological hyperperfused lesions.[3]

Interestingly, we found a non-monoexponential behavior of the simulated signal as a function of b-value, with a steep decline of the IVIM signal at very low b-value. This correlates well with our in-vivo experience, and steep declines at low b-value were already reported, see for example Fig. 6 (i) and Fig. 7 (n) in.[3] This could also be linked with the recent observation that a two pool model seems to better describe the IVIM cerebral perfusion in the rat.[21] The question arises why such a steeper decline is not always observed in in-vivo measurements. This might possibly be due to dephasing effects by the imaging gradients or noise causing an underestimation of the real signal at b = 0 s/mm$^2$.

The largest part of the relative volume of the network consisted of the capillaries. The signal decay was strongly dependent on the specific type of vessel (capillary, arterioles/venules or pial vessels) and we found that an important portion of the IVIM perfusion signal is coming from the arteries and veins inside the voxel. Of importance, only the effects arising from motion inside a specific vessel type was studied here, the effects arising from the blood particles moving between the vessel types were not considered. Further, one should note that the sensitivity with respect to boundary conditions in the blood flow simulation is the highest in the pial vessels. However, while this might affect the absolute velocity values, the impact of the pressure boundary conditions on the velocity ratio between the different vessels types is likely very small. In summary, our findings suggest that the IVIM perfusion signal does arise from all components of the microvasculature, not only the capillary bed as suggested by early theoretical assumptions.[22]

In conclusion, this simulation improves the theoretical understanding of the IVIM method, by directly linking the MR IVIM signal to ultra-high resolution measurements of the microvascular network and realistic blood flow simulation. The simulated pseudo-diffusion coefficient D* was found in the upper range of corresponding in-vivo measurements. In light of our results, not only the capillary bed, but all anatomical components of the microvascular network contribute to the MR IVIM perfusion signal.

|               | Network 1 |                  | Network 2 |                  | Network 3 |                  |
|---------------|-----------|------------------|-----------|------------------|-----------|------------------|
|               | D*        | Relative Volume  | D*        | Relative Volume  | D*        | Relative Volume  |
| **Whole network** | 31.7  |                  | 40.4      |                  | 33.4      |                  |
| **Artery**    | 284.9     | 10.7             | 172.1     | 20.1             | 833.3     | 9.6              |
| **Vein**      | 61.9      | 18.7             | 98.7      | 9.3              | 219.3     | 18.4             |
| **Arteriole** | 21.9      | 18.8             | 40.4      | 17.4             | 21.7      | 13.5             |
| **Venule**    | 60.5      | 11.7             | 38.8      | 19.1             | 27.1      | 24.8             |
| **Capillary** | 9.8       | 40.1             | 12.7      | 34.1             | 6.3       | 33.7             |

**Table 1**: Decomposition of the source of IVIM perfusion signal as function of the type of vessel, with D* [$10^{-3}$ mm$^2$/s] as obtained from a mono-exponential fit for the simulated MR signal decay for b-values (0-800 s/mm$^2$) and the relative volume of the whole network volume [%] of each vessel type. This table shows that an important portion of the IVIM perfusion signal originates from the arteries and veins inside the voxel.

Figures

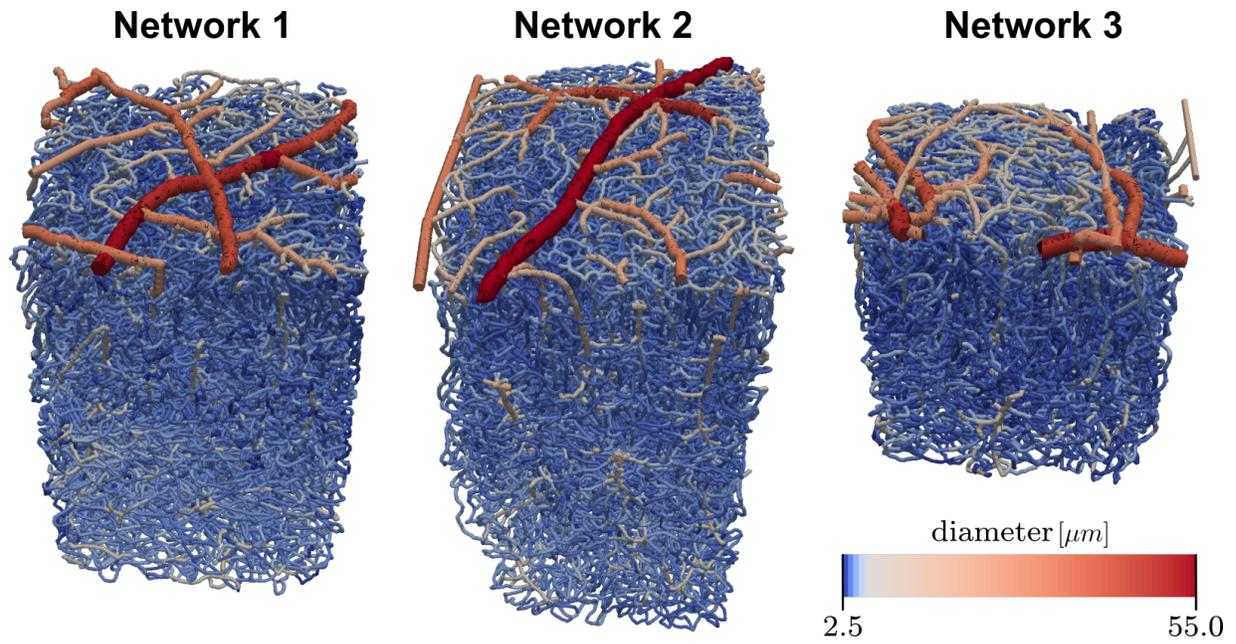

Figure 1. Three-dimensional visualization of the three microvascular networks. The vessels are colored based on the vessel diameters. The vessel segments are represented as tortuous tubes with constant diameter. For illustrative purposes the tube radii are enlarged.

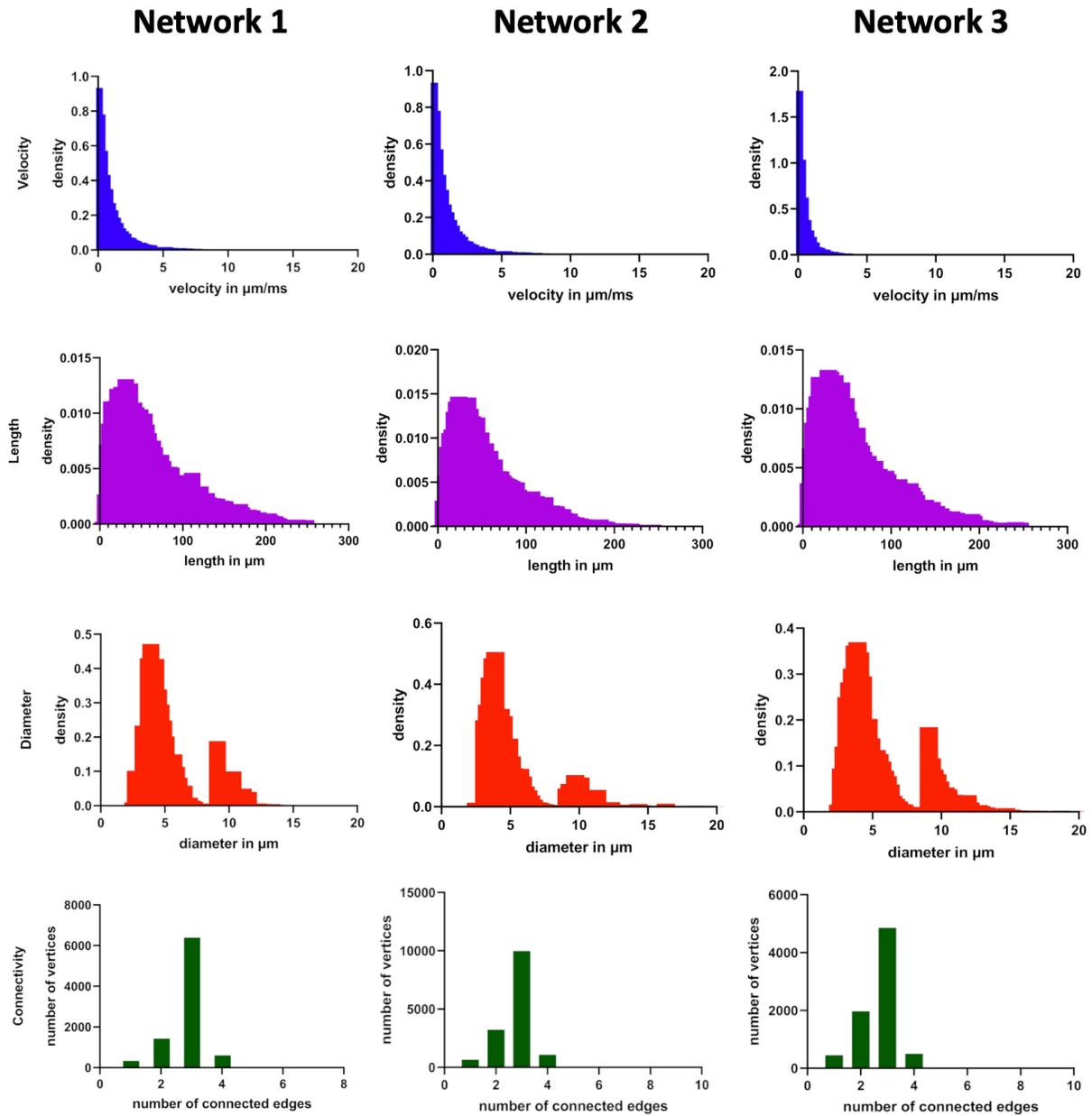

Figure 2. Velocity, vessel length, vessel diameter and connectivity distributions for all three networks.

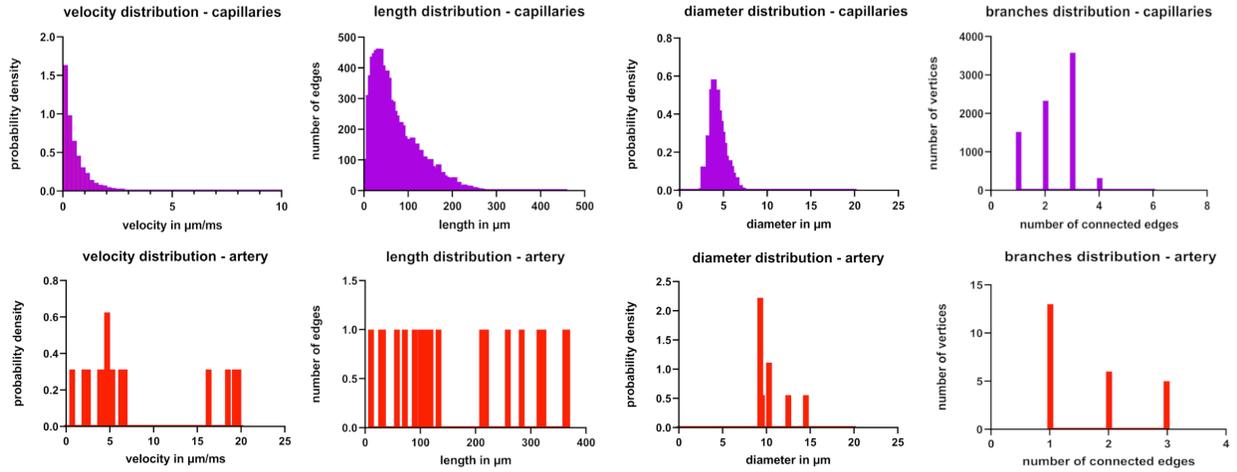

Figure 3. Capillaries and arteries velocity, vessel length, vessel diameter and connectivity distributions for all three networks.

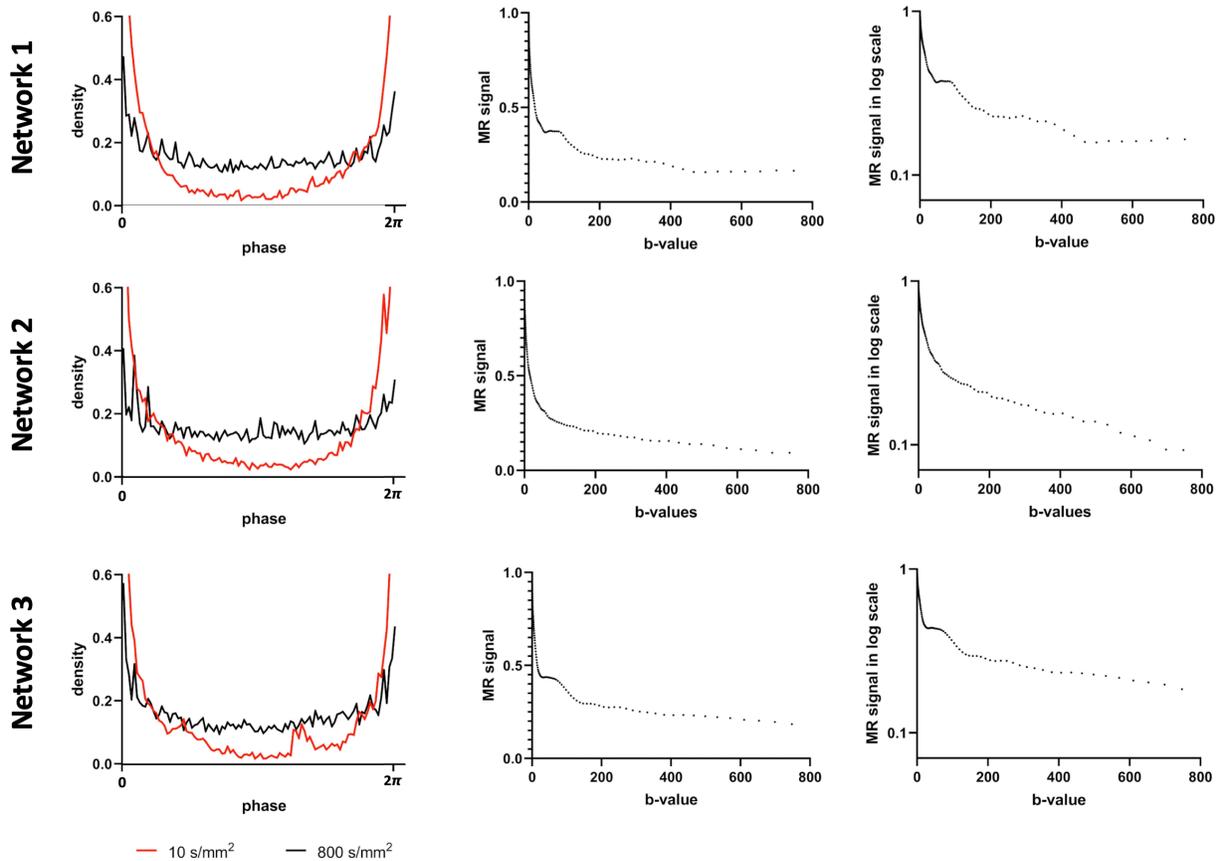

Figure 4. Phase and MR magnitude signal simulation for all three networks. **Left**: simulated phase distribution for b-values of 10 (*black curve*) and 800 s/mm$^2$ (*red curve*). **Middle**: simulated diffusion-weighted magnetic resonance magnitude as function of b-value. The fitted pseudo-diffusion coefficient D* were 31.7 x 10$^{-3}$ mm$^2$/s for network 1, 40.4 x 10$^{-3}$ mm$^2$/s mm$^2$/s for network 2 and 33.4 x 10$^{-3}$ mm$^2$/s mm$^2$/s for network 3, respectively. **Right:** simulated diffusion-weighted magnetic resonance magnitude as function of b-value in log-scale, showing the non-exponential behavior of the signal decay at low b-value. The phase is given in radian and the b-value in s/mm$^2$.

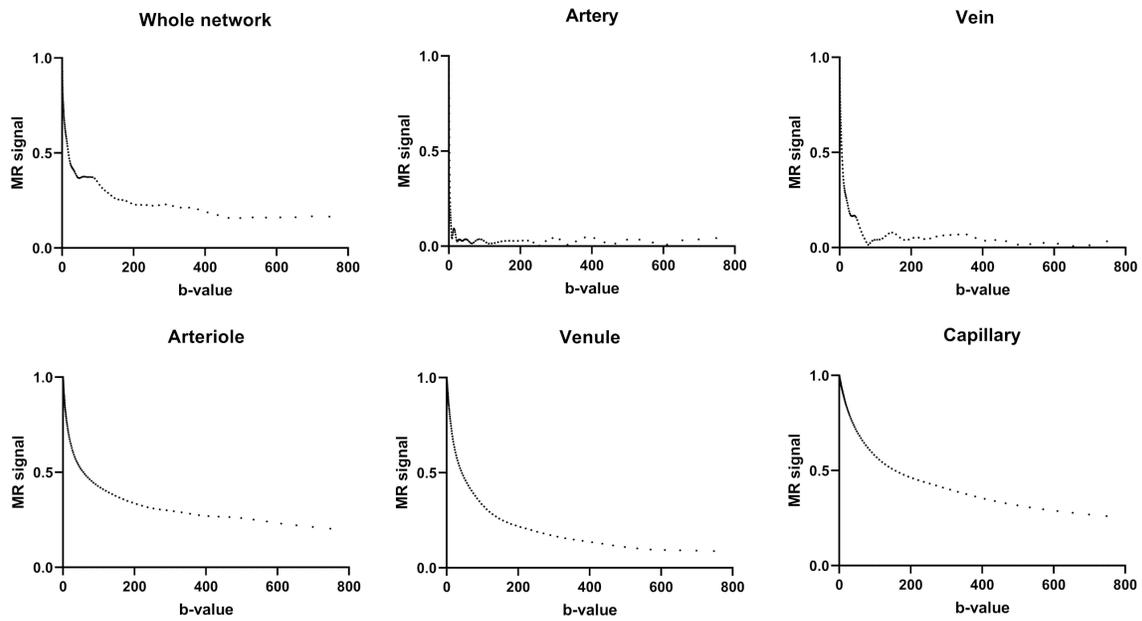

Figure 5. Decomposition of the source of the diffusion-weighted magnitude signal from specific vessel types for network 2: For low b-values, signal decay was the fastest in the artery network, while the signal decay from the arterioles and venules was slower, and the capillary had the slowest signal decay.

**AKNOWLEDGEMENTS**

Christian Federau was supported by the Swiss National Science Foundation (Grant No PZ00P3_173952 and CRSK-3_190697). Franca Schmid received funding from the European Union's Horizon 2020 Framework Program for Research and Innovation (Specific Grant Agreement No. 720270 [Human Brain Project SGA1] and Specific Grant Agreement No. 785907 [Human Brain Project SGA2]) and the Forschungskredit of the University of Zurich (grant no. FK-19-045).